\documentclass{kluwer}    % Specifies the document style.
\usepackage{epsfig}
\newdisplay{guess}{Conjecture}

\begin{document}                                                                                   
\begin{article}
\begin{opening}         
\title{Galactic Fountains and Galactic Winds%
%\dedication{Paper dedicated to the memory of Franz Kahn.}
} 
\author{Dieter \surname{Breitschwerdt$^{1,2}$} and 
Stefanie \surname{Komossa}$^2$}  
\runningauthor{D. Breitschwerdt and S. Komossa}
\runningtitle{Fountains and Winds}
\institute{$^1$Heisenberg Fellow\\
$^2$Max-Planck-Institut f\"ur Extraterrestrische Physik, 
Postfach 1603, D-85740 Garching, Germany, {\bf Email:} 
{\tt breitsch@mpe.mpg.de, skomossa@mpe.mpg.de}}

%\date{April 15, 1993}

\begin{abstract}
The development of galactic fountain theories is reviewed with special 
emphasis on the different approaches and concepts that have been used 
in the past. In particular the outstanding contribution of Franz Kahn to our 
{\em physical} understanding is appreciated. It is argued that galactic 
outflows represent an essential phase during galaxy evolution. The dynamics 
of the outflow imprints its signature on the emission spectra of soft X-rays, 
which may well be observable with AXAF and XMM. Finally, some remarks about 
winds in starburst galaxies and AGN are made. 
\end{abstract}
\end{opening}           
\section{Introduction}  
We present our review of galactic fountains and winds by emphasizing the 
different starting points and concepts that have led to its description. This 
stresses the underlying {\em physical ideas} and {\em pictures}, but is 
accomplished at the expense of not giving due credit to {\em all the 
individuals} that have contributed so much over the years, and to whom we 
apologize. 
\section{Overview of Fountains}
\subsection{The hot-phase approach}
\label{him1}
To our knowledge Shapiro \& Field (1976, henceforth SF76) were the first to 
coin the 
term ``galactic fountain''. It is interesting to note, that the idea 
developed in an effort to explain two key observations of a new hot 
interstellar phase as two different phenomena of the same plasma,  
viz.\ the ubiquitous interstellar O{\sc vi} line detected with 
Copernicus (Jenkins \& Meloy, 1974) and the 1/4 keV soft 
X-ray background (SXRB; Williamson et al., 1974). 
An outcome of these and subsequent investigations, however, was that these two 
observations should sample different regions in the interstellar medium (ISM). 
Such a conclusion is 
based on the assumption of collisional ionization equilibrium (CIE), and 
as we shall see, need not be as stringent as hitherto thought.
SF76 investigated two cases: a steady state and a 
time-dependent cooling model. They 
used the O{\sc vi} density $\langle n(OVI) \rangle$ as inferred 
from Copernicus, assuming CIE; the fractional ionization 
$n(OVI)/n(OI)$ is fixed for each temperature $T$. Likewise,  
the emissivity $\Lambda(T)$ in a given energy band is constant, and the 
SXRB intensity in the galactic plane is $\langle n_e^2 \rangle 
\Lambda(T)$. Apart from $T$, the only free parameter is the volume filling 
factor $f_{\rm V}$ of the hot component, which together generate a set of 
curves for the 
pressure $p/k_B = 2 n_e \, T$. For all admissible values of $f_{\rm V} < 1$, 
$p/k_B \geq 10^{4.2} \, {\rm cm}^{-3} \, {\rm K}$, which is obviously in 
contradiction with the O{\sc vi} line widths, which indicate that a substantial 
fraction of the gas is {\em below} $10^6$ K. 
A possible solution of this problem is that the assumption of steady-state,  
which enforces a global pressure equilibrium, is not valid. 
For if it were true, then a very efficient cooling mechanism of the shock heated 
plasma should exist. For an adopted value of $T = 10^6$ 
K, the model requires $n_e = 0.01 \, {\rm cm}^{-3}$ and $f_{\rm V} =0.4$ from 
which a cooling time of $10^7$ years results, an order of magnitude larger, 
as it were, than the reheating time of a gas element of $10^6$ years 
by supernova remnants (SNRs) (Cox \& Smith, 1974). Heat conduction in the 
ISM is also not efficient 
enough, and in our view in general overrated in its importance due to the 
efficient reduction of the mean free path of thermal electrons by even 
very modest magnetic field strengths - a point that was always emphasized 
by Franz Kahn. This leaves convection as the most efficient way of removing 
thermal energy from the disk. 

The previous arguments, although essentially correct, are largely model 
dependent (e.g.\ on the filling factor for the hot gas), and one can strengthen 
the argument for convection, following the analysis first given by Franz Kahn 
(1981). 
Owing to the fact that the hot intercloud medium (HIM), as deduced from the 
O{\sc vi} absorption line data, must have a scale height largely exceeding 
that of the warm and cool medium, one can ask if such a hot galactic 
corona\footnote{The existence of a galactic corona was postulated by Spitzer 
already in 1956 as a confining medium for high velocity clouds.} 
can be in hydrostatic equilibrium. The gas will settle down in equilibrium if 
the dynamical relaxation time scale, $\tau_{\rm dyn}$, is less than the cooling 
time scale, $\tau_{\rm cool}$.   

Defining the adiabatic parameter by $\kappa = P/\rho^{5/3}$, Kahn's (1976) 
cooling law\footnote{Franz Kahn, who combined physical insight with 
mathematical elegance discovered an analytical expression for the 
intrinsically complicated interstellar cooling law. Between a few $10^5$ to 
$10^7$ K, $\Lambda(T) \propto T^{-1/2}$, which results in an integrating 
factor for the time-dependent energy equation.} 
states: $D\kappa^{3/2}/Dt = -q$,
with $q = 4 \times 10^{32} \, {\rm cm}^6 {\rm g}^{-1} {\rm s}^{-4}$, and thus 
$\tau_{\rm cool} \simeq \kappa^{3/2}/q$.
For $\tau_{\rm dyn} \ll \tau_{\rm cool}$,  
$c_h^2/\rho_h \gg \gamma^{3/2} q/g_z$, 
with $g_z \sim 5 \times 10^{-9} \, {\rm cm}/{\rm s}^2$, being the 
$z$-component of the gravitational acceleration towards the disk, and $c_h$ 
and $\gamma$ denote the speed of sound in the hot medium and the ratio of 
specific heats, respectively. For a pressure and density of the corona of 
 $P_h \sim 10^{-12} \, {\rm dyne}/{\rm cm}^{2}$ and $\rho_h \sim 10^{-26} \, 
{\rm g}/{\rm cm}^{3}$, the above inequality is heavily violated 
by one order of magnitude. Does that 
rule out hydrostatic halo models once and for all? To answer this, 
the equilibrium condition is rewritten as $P_h/\rho_h^2 \gg \gamma^{1/2} q/g_z$ 
or  $c_h^4/P_h \gg \gamma^{5/2} q/g_z$. 
The RHS does only vary with galactic radius $R_0$, since $g_z \propto z/R_0^2$ 
(Kahn, 1981), provided that $z/R_0 \ll 1$. Thus for a fixed equilibrium 
pressure $P_h$, the outcome depends 
heavily on $c_h$, and therefore on the respective values of $\rho_h$ and $T_h$.    
In other words, although for ``standard'' HIM values the halo should be 
{\em dynamic}, hydrostatic solutions may be possible for high temperatures 
and/or low densities. Physically this means that the HIM is too hot to be 
confined by the gravitational field to a thin layer, rising into the halo 
towards a ``thick disk''. However, radiative cooling is too strong in order 
to attain an equilibrium configuration, unless there is heavy depletion in 
metals. 

A crude picture of fountain dynamics was given by SF76, 
and considerably refined and improved by Kahn (1981).  
The uprising gas will move subsonically a distance $z_c \sim v_z 
\tau_{\rm cool}$, with $v_z \le c_h$, or $z_c \le c_h^6 /(\gamma^{5/2} q P_h)  
\sim 1$ kpc for $c_h=130$ km/s. At height $z_c$ thermal instabilities  
will promote cloud formation. The clouds travel ballistically until their 
kinetic energy is used up at roughly a distance $z_{\rm max} \sim v_z^2/2 g_z$ 
$\approx 5.5$ kpc, and gravity pulls them back to the 
disk, which in a 1D fountain is the point of origin of the hot gas. 
Disregarding halo magnetic fields, there is no drag on the clouds and they 
will strike the disk at a velocity $-c_h$, and may tentatively 
be identified with the so-called high velocity clouds\footnote{For practical 
purposes HVCs are defined as H{\sc i} clouds that deviate more than 50 km/s
from the velocity range allowed by a simple model of galactic rotation 
(Wakker, 1991). Note that a predominance of negative velocities is observed.} 
(HVCs), studied in absorption towards background stars (s.~Wakker \& van
Woerden, 1997). 
The mass flux of hot gas into the halo is roughly $\dot M = 2 \pi R_g^2 
f_{\rm V} \rho_h v_z \approx 29 \, f_{\rm V}\, {\rm M}_\odot/{\rm yr}$. 
SF76 used slightly different HIM values, i.e.\ $n_h = 0.01 \, {\rm cm}^{-3}$ 
and $c_h = 170$ km/s , which yields $\dot M \sim 30  {\rm M}_\odot/{\rm yr}$ 
for $f_{\rm V} \sim 0.4$, in rough agreement with the observationally 
detected HVC flux. 

O{\sc vi} absorption line widths show (Jenkins \& Meloy, 1974) that almost 
one third of the gas has temperatures below $5 \times 10^5$ K. 
In the frame of the steady-state model SF76 could show that $P_h/k_B \ge 
10^6$, implying a number density $n_h > 1 \, {\rm cm}^{-3}$. In fact, 
it turns out that for values of $n_h > 0.1 \, {\rm cm}^{-3}$ the 
radiative cooling time is less than the SNR reheating time. Therefore
Shapiro \& Moore (1976) calculated time-dependent cooling models.
It is concluded that subsequent 
adiabatic expansion would not allow the plasma to spend enough time in 
the energy range of 1/4 keV X-ray emission. Spectra of isochoric cooling 
are presented only down to a temperature of $10^{5.6}$ K. However, 
more recent calculations of the SXRB, including the dynamical effects of 
expansion and time-dependent cooling in a self-consistent fashion, show that 
the emergent X-ray spectrum down to a temperature of 
$\sim 4 \times 10^4$ K is still reminiscent of a $10^6$ K gas, copiously 
emitting photons at 1/4 keV by delayed recombination (Breitschwerdt \& 
Schmutzler, 1994). We will come back to this point below.

\subsection{The HVC approach}
A completely different approach was taken by Bregman (1980), who analyzed 
the HVC flow and simulated it by a 2D axisymmetric hydrodynamical model. 
Although distances of HVCs are poorly known in general, they appear as 
prominent chains on the sky covering a fraction of more than 10\%. As the 
observed HVC mass flux is in excess of $1 \, {\rm M}_\odot/{\rm yr}$, the 
integrated HVC mass over a Hubble time is larger 
than the present disk mass. The most likely explanations are that (i) if HVCs 
are accreted, the process is intermittent or (ii) if the source of the HVCs is 
the disk gas itself, then it enters a cycle of upwards motion, cooling, 
condensing and raining back onto the disk. The main ingredients of the model 
are (i) a realistic gravitational potential, consisting of a bulge, disk and a 
spherically symmetric dark matter halo component, (ii) galactic 
rotation and therefore conservation of angular momentum, and (iii) equilibrium 
cooling. In particular rotation has some  interesting effects, notably
that the disk and halo are not in corotation, for the following reason.
Let $\Omega(R_0)$ be the position dependent rotational frequency of the 
differentially rotating galactic disk, 
so that a gas element has specific angular momentum $l_0=\Omega(R_0) R_0^2$.  
Centrifugal force balance, $\Omega^2 r = g_{\rm eff}$, and conservation of 
angular momentum require $r=\left(\Omega^2(R_0) R_0^4/g_{\rm eff}(z) 
\right)^{1/3}$, where $g_{\rm eff}$ is the gravitational acceleration now 
depending on $r$ and $z$. Since $g_{\rm eff}$ decreases with height $z$, 
the uprising gas will also move radially outwards. It has been shown 
however, that inclusion of a magnetic field, anchored in the  
gaseous disk, can enforce corotation of the halo due to magnetic stresses up 
to the Alfv\'enic critical point, typically at a vertical distance of a 
few kiloparsecs (Zirakashvili et al., 1996).

Additional radial motion in Bregman's model is generated by a radial pressure 
gradient in the disk, inferred from pressure equilibrium between the hot, warm 
and cold ISM phases, $P_h(r) \propto n_{\rm HI}(r)+n_{\rm H2}(r)$, decreasing 
with galactocentric radius beyond 6 kpc. Once clouds have formed they lose 
pressure support and move ballistically. 
In summary, the favoured model takes as input a density and temperature at 
the base of the corona of $n = 10^{-3} \, {\rm cm}^{-3}$ and 
$T = 10^6 \, {\rm K}$. According to the simulations, the gas travels 2-3 times 
its original radial position and $5 - 10$ kpc in height before clouds are 
formed, returning to the disk $3 - 60 \times 10^7$ yr thereafter. The 
mass flux onto the disk is $2.4 \, {\rm M}_\odot/{\rm yr}$ and the total mass 
of the corona is $\sim 7 \times 10^7 \, {\rm M}_\odot$. The crude distribution 
of HVCs, in particular a chain-like appearance, can be reproduced.

There are other interpretations of the HVC phenomenon 
than the galactic fountain, and at least two more processes are needed for 
explaining the Outer Arm extension and the Magellanic stream (for a detailed 
discussion see Wakker \& van Woerden, 1997). 
\subsection{The analytic approach}
According to Francis Crick, ``A theory that agrees with all the data at 
any given time is necessarily wrong, as at any given time not all the data 
are correct.'' There lies the advantage of an analytical fountain model, 
because it concentrates on only the most important observational facts 
-- subject to personal judgment --
in order to gain a physical understanding of the basic processes involved. 
Franz Kahn was a master in this respect, and his seminal paper (Kahn, 1981) 
shows this impressively. He must have certainly disliked ``cartoon physics'', 
as there is no single figure within 28~pages (two tables is all the superficial 
reader can hope for), but instead everything 
is developed mathematically {\em ab ovo}. First he derives a useful 
approximation for the gravitational potential, consisting of a flat disk and 
a spherical halo distribution, ascribing the latter to Population~II objects; 
interestingly, dark matter is not mentioned. Next he analyses the cooling of 
the hot gas streaming away from the disk, and shows that for a wide and most 
likely range of HIM parameters, such as $\rho_h$ and $T_h$, a hydrostatic 
equilibrium does not exist (cf.\ Section~\ref{him1}). In the next step 
the heat input by supernovae (SNe) is discussed, assuming that SNe occur 
randomly over the whole disk and the ambient medium is cool and has 
uniform density $\rho_0$. By overestimating $\rho_0$, the effect of HIM 
heating is underestimated, since the adiabatic parameter $\kappa$ increases 
with decreasing density, and it is the gas with the highest values of 
$\kappa$ which takes longest to cool. 
Heat conduction being unimportant, 
an element of fluid retains its value $\kappa_s$ once it was shocked, 
until cooling becomes severe. 

For illustration, it is shown how neatly one can calculate the mass of hot gas 
contributed to the HIM by a single SNR.
Starting out with the Sedov solution for SNRs, $r_s=a t^{2/5}$ with 
$a=(2 E_{\rm SN}/\rho_0)^{1/5}$, and using strong shock conditions 
$P_s = (3/4) \rho_0 \dot r^2$ and $\rho_s = 4 \rho_0$, where $P_s$ and 
$\rho_s$ are the post-shock pressure and density, respectively, we find that 
$\kappa_s \propto t^{-6/5}$. The accumulated mass behind the shock is 
$M_s = (4/3) \pi \rho_0 r_s^3 \propto t^{6/5}$, and therefore we obtain 
the important result that $\kappa_s M_s = const. = 
0.1 E_{\rm SN}/\rho_0^{2/3}$. The remnant cools fastest in the outer parts, 
since the post-shock temperature $T_s$ is lower in the more recently shocked 
gas, because the shock speed drops as $\dot r_s \propto t^{-3/5}$, and 
$T_s \propto \dot r_s^2 \propto t^{-6/5}$. So when radiative cooling sets in, 
there are still parts of the remnant which have $\kappa_s > \kappa_*$, and which 
retain hot material. The mass that an individual remnant roughly contributes 
to the HIM can be evaluated in the following way. 
$d M_s = \rho_s dV_s = - 0.1 E_{\rm SN} d\kappa_*/(\kappa_*^2 \rho_0^{2/3})$, 
from which by integration follows for the volume occupied by the mass 
$M_* = M(\kappa_s > \kappa_*)$: $V_s = (5/2) M_* \kappa_*^{3/5} P_s^{-3/5}$, 
and thus the smoothed out density of the still hot gas is $\rho = M_*/V_s = 
(2/5) P_s^{3/5} \kappa_*^{-3/5}$. For $P_h = 10^{-12}$ and $\rho_h = 10^{-26}$,
 one obtains $\kappa_* = 4.7 \times 10^{30}$ (all in CGS units).Therefore the 
mass per SN contributed to the HIM is $M_* = 0.1 E_{\rm SN}/(\rho_0^{2/3} 
\kappa_*) \approx 320 \, {\rm M}_\odot$, for $E_{\rm SN}=3 \times 10^{51}$ erg. 
There is one further assumption that enters here: the calculation only holds 
as long as different parts of the gas do not mix; however at some stage the 
hot gas becomes part of the HIM, and it is assumed that mixing will then occur 
instantaneously, with an energy injection of $E_0 = 5 P/(2 \rho) \approx 2.5 
\times 10^{14} \, {\rm erg}/{\rm g}$. The cooling time for this gas is about 
$8.3 \times 10^6$ years, much larger than the time needed for a SNR to come 
into pressure 
equilibrium with the HIM, and therefore the gas 
must expand into the halo. For a galactic SN rate of $1/30 \, {\rm
yr}^{-1}$, the mass flux into the fountain on either side is $5.3 \, 
{\rm M}_\odot/{\rm yr}$. 

Next we have a look at the fountain flow. For lack of space, we will restrict 
ourselves to a qualitative discussion. Clearly, in a dynamical halo radial 
motion of the gas is suppressed, if the vertical dominates the radial pressure 
gradient. The fountain described in the following is strictly one-dimensional.
Physically the flow is controlled by the processes of energy injection at the 
base, radiative cooling and gravitational pull. In contrast to Bregman's (1980) 
fountain, the cooling time is ten times shorter due to the higher initial 
density, and therefore the characteristic cooling length is much shorter than 
the maximum height of the fountain. Once cooling occurs energy losses become 
catastrophic, and one can divide the flow into two distinct regions: in the 
lower fountain cooling dominates and gravity is unimportant, and vice versa 
in the upper fountain, where the gas moves ballistically. 

In the lower part, conservation of mass, momentum and energy of a steady 
flow, ensures the existence of a critical layer, where the flow becomes 
transsonic. For $10^{14} < E_0 < 2.5 \times 10^{14} \, {\rm erg/g}$ 
the sonic point is located at $160 < z_c < 440 \, {\rm pc}$. The gas 
keeps on losing thermal energy beyond $z_c$ and at a distance of 
$60 < z_{\rm cool} < 450 \, {\rm pc}$ from there the flow becomes pressureless, 
and reaches terminal velocity $u_{\infty}$. In the upper part, at roughly 1 kpc 
above the galactic plane, gravity determines the maximum height  
by the implicit equation $u_{\infty}^2 = 2 \Phi_0(z_{\rm max})$, which gives 
$1570 < z_{\rm max} < 5093 \, {\rm pc}$; thus the ballistic region is about 
six times as large as the cooling region. The rise time to $z_{\rm max}$ is 
$t_{\rm rise} \sim \int_0^{z_{\rm max}} dz/u$ or $3.8 \times 10^7 < 
t_{\rm rise} < 5\times 10^7$ yrs.

Once the flow reverses in a 1D-model it has to interact with the upward flow. 
The gas in the descending sheet is so cool that its sound speed is much 
lower than the downward flow velocity $u_{\rm down}$. Applying the 
equation of motion and conservation of mass and momentum, it can be deduced 
that $u_{\rm down} = 1/2 u_{\rm up}$. The gas moving upwards 
joins the sheet after passing through a shock, leading to a reduced effective 
gravitational acceleration $g_{\rm eff} = (3/4)\, g_z$. It can be shown that 
the cooling of the shock heated gas is so fast that gravity does not have any 
effect on its structure. The thickness of the cooling layer is 
$2\, {\rm pc} < \Delta z_{\rm cool} < 385 \, {\rm pc}$, which is indeed very
thin. 

Is the flow just described stable against linear perturbations? It is easy 
to show that in the plane parallel case no growing modes exist. This can 
be understood as follows: moving the sheet ahead of its zero-order 
location increases gravity, but this is over-compensated by the momentum 
pick-up from the ascending flow, so that as a whole the sheet is decelerated. 
However, here is a serious draw-back of the 1D-calculation. In a 2D-case, 
it is very likely that the descending sheet will become Rayleigh-Taylor 
unstable due to the rising gas from below, which now does not need to pass 
through an orderly shock. Instead, the sheet can fragment, 
and the uprising gas will rather interact with downwards moving {\em clouds}. 
On the other hand, if a disk-parallel magnetic field is present in the 
halo, then it would have a stabilizing effect. In addition the cooling layer 
will be puffed up by magnetic pressure support, becoming thicker than the 
radiating layer. Whereas small scale disturbances can grow to large 
amplitude they would only cause a corrugation of the bottom surface of the 
sheet, and disturbances with wavelengths of the order of the sheet would 
be prevented from growing, because the distorted shock would stabilize it. 
Thus at the level of this analysis, the sheet structure is disturbed but 
not destroyed. 

In subsequent papers, Franz Kahn investigated the effects of the magnetic field 
in more detail, in particular the possibility of a fountain dynamo 
(Kahn \& Brett, 1993). The most important improvements, however,  
he worked out shortly before his death (Kahn, 1998), when he described, again 
analytically, SN explosions in a hot rarefied medium like in superbubbles.
He shows that subsequent off-centre explosions can still be described by 
concentric similarity solutions, because the shock wave advances rapidly 
through the rarefied medium and spends most of the time propagating through 
the outer shell, where the bulk of the mass sits. The dynamical effect on the 
ISM is largest, if SNRs expand in the HIM, because radiative cooling is 
less important than adiabatic cooling. These considerations led to the 
3D fountain calculations of Avillez (1998), in which it is shown, that once 
the ISM is disturbed by explosions it will never return to the original state, 
but after some time a frothy disk of half-thickness 1 kpc will develop. 
Mixing with cooler gas generates shear and turbulence and the fountain is far 
from smooth. These calculations represent an important step forward, but need 
to be pursued further to include the effects of magnetic fields 
in conjunction with grid refinement (in order to resolve small scale structures 
that are present), and above all to include non-equilibrium cooling. 
The importance of the latter will be discussed in the next section. 

\section{Galactic Winds}
\subsection{Winds in spiral galaxies}
It is generally believed that galactic winds in normal spiral galaxies do not 
play a significant r\^ole during the galactic evolution. The argument is 
based on the high temperature in the disk-halo interface needed for thermal 
winds; e.g.\ Habe \& Ikeuchi (1980) estimate that $T > 5 \times 10^6$ K 
near the solar circle, if $n = 10^{-3} \, {\rm cm}^{-3}$. However, this 
completely ignores the dynamical effects of the non-thermal component of the 
ISM, i.e.\ the cosmic rays (CRs), which have a local energy density comparable 
to the thermal plasma. 
Moreover, CRs are coupled to the thermal gas, via frozen-in magnetic field 
lines, having a gyroradius of 
only $r_g = 3.3 \times 10^{12} \, R_e[GV]/B[\mu{\rm G}]\, {\rm cm}$, with 
$R_e= p c/Z e$ being the rigidity ($p$ and $Z$ are the momentum and nuclear 
charge of the particle, respectively). The effective propagation 
speed of the CRs through the ISM is drastically reduced from the speed of 
light, $c$, to the Alfv\'en speed, $v_{\rm A} \sim 80$ km/s in the HIM, because 
CRs strongly interact with magnetic field fluctuations $\delta B$. In 
particular, if the wavelength of perturbations is of the order of the
gyroradius, CRs suffer strong scattering in pitch angle and perform a random 
walk through the plasma, usually described by a diffusion process. It is known 
from the measurement of secondary CR unstable isotopes like $^{10}$Be, that 
the bulk of CRs leaves the Galaxy after about $2\times 10^7$ yrs, and therefore 
there exists a spatial gradient in the CR distribution function. It has been 
shown that a small scale anisotropy of pitch angle distribution in the 
wave frame results in a so-called streaming instability, 
leading to a resonant generation of waves in order to remove the cause 
of the instability. McKenzie \& V\"olk (1982) have shown that this effect 
can be described in a hydrodynamic fashion by a CR pressure gradient, $\nabla 
P_c$. The overall effect is a net momentum transfer from CRs to the gas via 
waves as a mediator, pushing away the gas from the disk into the halo.  
It has been demonstrated that in a flux tube model, including a realistic 
bulge/disk 
and dark matter halo gravitational potential, this process can lead to 
a secular mass loss of gas (along with CRs) of the order of 
$\dot M_{\rm GW} \sim 0.5 \,{\rm M}_\odot/{\rm yr}$, given an initial density 
of $n_0 = 10^{-3} \, {\rm cm}^{-3}$, $T_0= 10^6 \, {\rm K}$, $P_{\rm c0} = 
10^{-13} \, {\rm dyne}/{\rm cm}^2$ (Breitschwerdt et al., 1991). To be a 
bit more specific, it is assumed that there exists a magnetic field component
perpendicular to the disk, like e.g. in NGC$\,$4631, 
with $B_{\perp} = 1\, \mu{\rm G}$, 
bounding the flux tubes. In order to ensure strong coupling between CRs and 
gas, the level of wave turbulence $\delta B/B$ can be quite small; in fact, 
non-linear Landau damping can limit wave growth, leading to dissipative 
plasma heating (Zirakashvili et al., 1996; Ptuskin et al., 1997). 

\begin{figure}
 \centerline{
  \epsfig{file=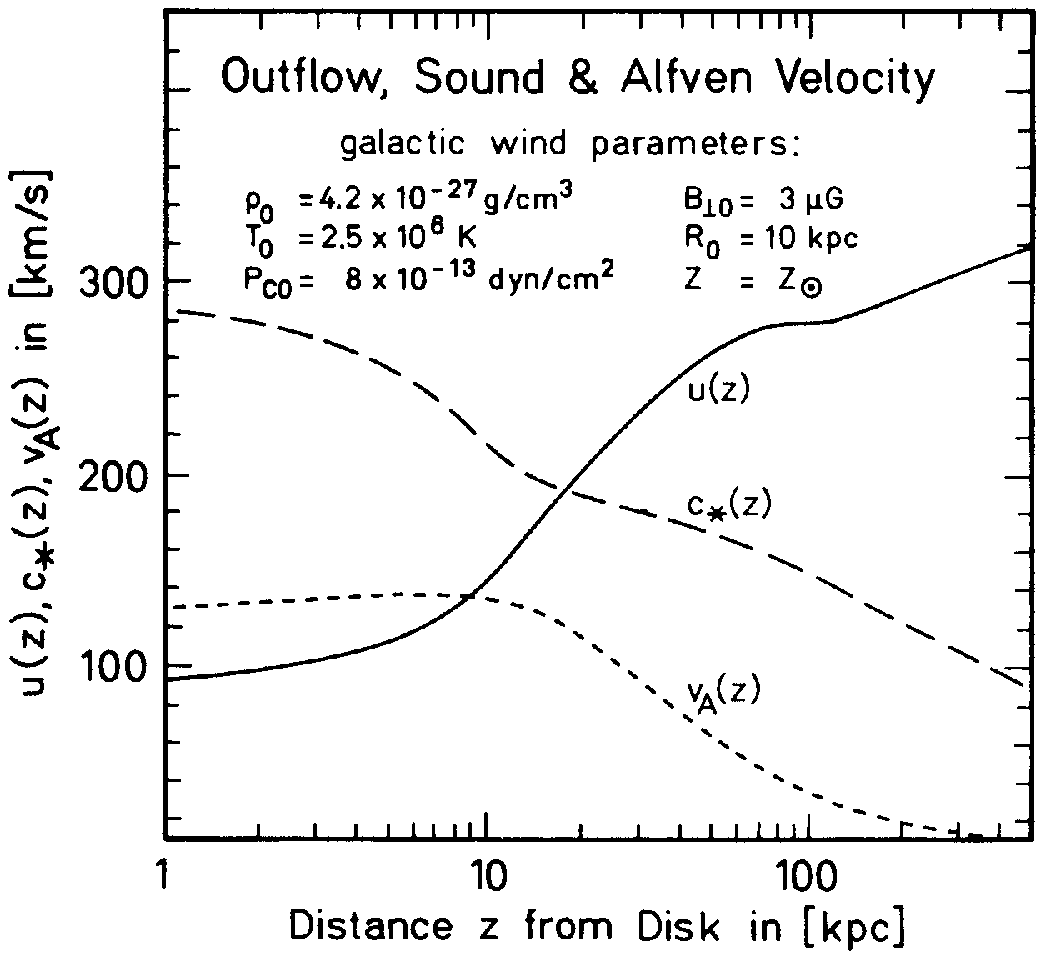,width=0.5\hsize,clip=}
 %    \quad
  \epsfig{file=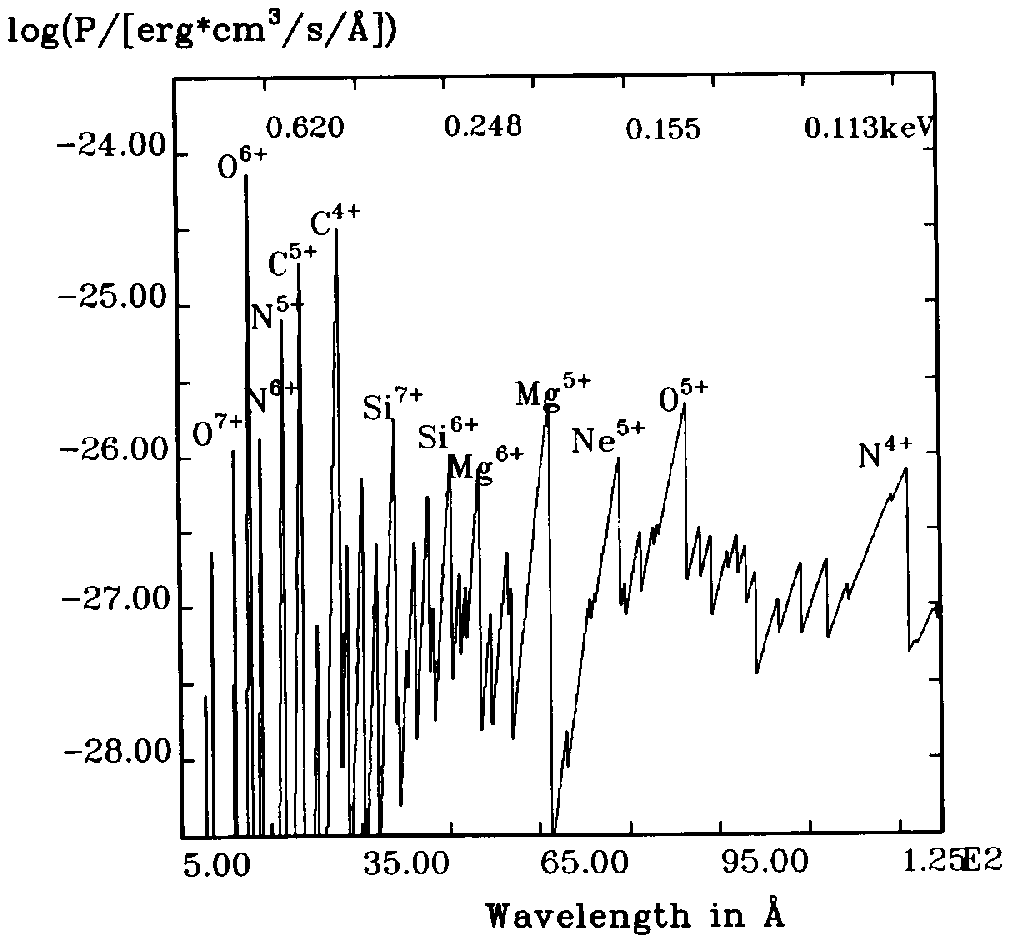,width=0.5\hsize,clip=}
            }
    \caption[]{{\em Left:} {\bf a)} Outflow velocity $u(z)$, Alfv\'en velocity 
        $v_{\rm A}(z)$ 
        and ``compound sound speed'' $c_*(z)$ (accounting for thermal, CR 
	and wave disturbances) for a {\it local} galactic wind, with boundary 
	conditions indicated.               
        {\em Right:} {\bf b)} High resolution photon spectrum for the local 
	non-equilibrium 
	emission from a local galactic wind, normalized to electron density 
	${n_{\rm e}}^2$ at a kinetic temperature of $4.1 \times 10^4$ K.   
	       }
    \label{Fig-lgw}
   \end{figure}

As it turns out in the steady state model, the combined pressure forces of gas 
and CRs can drive an outflow to infinity above a CR diffusion region of height 
$|z| \ge 1$ kpc, and there exists a 
subsonic-supersonic transition in the flow, like in Kahn's fountain,  
if the total pressure at the critical point exceeds the intergalactic pressure, 
$P_{\rm IGM}$. For finite $P_{\rm IGM}$ there must be a termination shock at 
a distance roughly given by $\rho_{\infty} u_{\infty}^2 \sim P_{\rm IGM}$, 
or $r_{\rm ts} \sim (u_{\infty} \dot M_{\rm GW}/P_{\rm IGM})^{1/2} \approx 
300 \, {\rm kpc}$, for $u_{\infty} = 300$ km/s and $P_{\rm IGM} = 10^{-15} \, 
{\rm dyne}/{\rm cm}^2$. In clusters of galaxies, $P_{\rm IGM}$ can be large 
enough to quench steady-state winds (V\"olk et al., 1996).
The importance of CRs is a consequence of their basically infinite scale 
height\footnote{Relativistic particles do not feel a gravitational field in 
the galaxy.} and that they are not subject to cooling\footnote{The 
dynamically important CRs are the nucleons; only electrons, which make up  
1\% in total, will lose energy by synchrotron emission and inverse Compton 
effect.} except adiabatic energy losses by $p dV$-work. Like in the case of 
the galactic fountain, the specific enthalpy of the gas (and the CRs) supplied 
at the base of the flow determines the structure of the wind. 
In spirals, we can crudely assume that there are {\em global} outflows, 
wherever hot gas percolates through the extended H{\sc i} and H{\sc ii} 
layers, and {\em local} outflows, driven by one or more 
superbubbles\footnote{It is often thought and shown in numerical simulations 
that a disk-parallel magnetic field can severely inhibit an outflow. However, 
such a field is susceptible to Parker instability (see Kamaya et al., 1996).}. 
A typical local outflow, which might produce part of the SXRB, is shown in 
Fig.~\ref{Fig-lgw}. Note that in Fig.~\ref{Fig-lgw}b a snapshot of the 
emissivity of the expanding wind is taken when the kinetic temperature is 
already down to $4.1 \times 10^4$ K, and most of the emission is due to 
{\em delayed recombination} (Breitschwerdt \& Schmutzler, 1994; 1999), because 
collisional excitation of lines in the soft X-rays is impossible.  
%
%----------------------------------------------------------- 
%
%______________________________________________________________
%  
Here another word of caution has to be added. In the literature one can 
frequently find hydrodynamical calculations including ``non-equilibrium'' 
cooling, like e.g.\ in Houck \& Bregman (1990) for a low-temperature 
galactic fountain, who give a ``non-equilibrium cooling function'' for a 
gas with cosmic abundances in the energy range $10^4 - 10^6$ K. It is argued 
that the error by ignoring the detailed ionization structure is small, because 
the total emitted power should be roughly constant. It is clear that if one is 
interested in the emergent spectrum, the distribution of ionization stages 
cannot be neglected. But more importantly, such an assumption ignores the 
dynamical and thermal coupling of the plasma. In an expanding plasma like a 
fountain or 
wind, the dynamical time scale can be shorter than the ones for ionization and 
recombination. Therefore the cooling function evolves in time and has to be 
calculated {\em self-consistently} with the dynamical and thermal structure of 
the plasma at each time-step (Breitschwerdt \& Schmutzler, 1999).     
\subsection{Winds in starburst galaxies and AGN}
The existence of galactic winds is most obvious in starburst galaxies, and 
for some reason these winds have been coined ``superwinds''. 
They are most easily detected in the optical and X-ray spectral
region. In X-rays they show up by their spatial extent (excellently traced
by the ROSAT HRI instrument; Tr\"umper 1983)
and characteristic emission line spectra (detected with ASCA), 
and many superwinds 
have been found in recent years in starburst galaxies
(e.g., NGC\,2146, NGC\,4945, NGC\,4666, Mrk\,1259, NGC\,2782, NGC\,1365, 
NGC\,3079) 
and mergers (e.g., Arp\,299, NGC\,6240).  
Prominent examples 
are M$\,$82 and NGC$\,$253, which have been reported to be underabundant 
in heavy elements (e.g.\ Ptak et al., 1997). It seems to us that this is an 
artifact and more likely a 
consequence of non-equilibrium cooling, which must be very strong due to the 
high flow speeds (e.g.\ $\sim 1000$ km/s in M$\,$82). Preliminary calculations 
have shown that indeed the collisional excitation of the Fe-K$\alpha$ line 
in a fast expanding wind is weak. 

In active galactic nuclei (AGN), winds are expected to develop in the central 
region. They may play an important r\^ole in AGN unification schemes
and are linked in some way or the other to the `central engine': 
Firstly, winds are expected to form in relation to accretion disks, 
removing angular momentum of the accreted matter.
Several mechanisms have been explored to drive the
wind, like centrifugal forces or radiation pressure 
(e.g., K\"onigl \& Kartje, 1994, Murray et al., 1997).
Secondly, near-nuclear dusty gas is very sensitive to radiation
pressure and strong winds will be driven this way (e.g., Pier \& Voit, 1995).
Components of the active nucleus that contain dust are 
the putative molecular torus and/or the so-called dusty `warm absorbers'
(see Komossa et al., these proceedings).
In particular, there are both, observational indications
(see, e.g., the case of NGC\,4051; Komossa, 1999) and theoretical
expectations (e.g., Krolik \& Kriss, 1995) 
that this warm material is not in ionization equilibrium, 
whereas this assumption is usually imposed when
modelling warm absorbers. 
Thirdly, on kiloparsec scales, outflows could originate when AGN-driven 
jets entrain and heat gas as they make their way out of the galaxy
(Colbert et al., 1998).
\section{Summary and Outlook}
Galactic outflows can be divided into fountains and winds, depending on 
whether some of the mass is lost in a secular fashion or not. Observationally, 
the difference is hard to tell by just looking at the rising gas, because one should 
be able to measure velocity and density {\em gradients}. But only if the 
flow were steady-state one would be able to discriminate. However, it is 
very likely that outflows are {\em time-dependent}, since a typical flow time 
of $10^8$ years is larger than the life time of superbubbles, 
the most efficient driving sources. Nonetheless, substantial progress will be 
made with high-resolution X-ray spectroscopy by AXAF and XMM, because 
the dynamical signature of hot outflowing gas will be present in the 
spectra due to the non-equilibrium ionization structure. Clearly, high-velocity 
clouds should be a consequence of fountain return flows. 

Fountains and winds are {\em not} mutually exclusive, but rather represent 
different {\em modes} of outflow, realized by different initial conditions 
in density, temperature, CR pressure and magnetic field strength and topology 
near the disk. If for some reason, CR coupling to the plasma is lost and 
CR diffusion dominates advection, a fountain will ensue (Breitschwerdt et 
al., 1993), coexisting with a wind flow nearby. 

Finally, outflows are an essential phase in galaxy evolution, because they 
prevent the reheating catastrophe of the disk and thus make star formation 
a continuous process. In particular winds take away a
significant fraction of chemically enriched material over a Hubble time. 
\vskip0.3cm
\noindent DB acknowledges financial support from the {\em Deutsche Forschungsgemeinschaft}
by a Heisenberg Fellowship.

\end{article}
\end{document}